\documentclass[preprint,12pt]{elsarticle}

\usepackage[a4paper, total={6in, 9.2in}]{geometry}
\usepackage[utf8]{inputenc}
\usepackage{mathabx}
\usepackage{graphicx}
\usepackage{siunitx}
\usepackage{hyperref}
\usepackage{adjustbox}
\usepackage{subcaption}
\usepackage{float}
\usepackage{placeins}
\usepackage{stfloats}
\usepackage{comment}

\journal{}

\biboptions{numbers}

\date{\today}

\begin{document}
\begin{frontmatter}
\title{Event-sparse stack denoising for 4D-STEM applications}

\author{Gregory Nordahl}
\author{Rebekka Klemmt}
\author{Espen Drath B{\o}jesen}
\ead{espen.bojesen@inano.au.dk}
\address{Center for Sustainable Energy Materials (CENSEMAT) \& Interdisciplinary Nanoscience Centre, Aarhus University, Denmark}

\begin{abstract}
We introduce a denoising method for four-dimensional scanning transmission electron microscopy (4D-STEM) that relies on processing local, scan position–independent electron event-sparse data stacks, called event-sparse stack denoising. This method adds an extra time dimension during data collection by recording multiple electron event–sparse diffraction patterns. The resulting datasets are effectively five-dimensional, referred to as locally time-resolved STEM (LTR-STEM). Diffraction data stacks at each scan position are processed using one of two sparsity denoising pipelines: 1) the density-based spatial clustering of applications with noise (DBSCAN) algorithm followed by multi-step persistence thresholding, or 2) sparse principal component analysis (sparse PCA), followed by single-step thresholding. Both methods perform well for diffraction data denoising, as shown by simulated peak signal-to-noise ratio (PSNR) curves, denoised experimental data for virtual imaging, and application-specific denoising for defect detection. PSNR analysis indicates that sparsity-denoised 4D-STEM data reaches the same PSNR as raw data at approximately \SI{16}{\percent} of the exposure time, demonstrating comparable image quality with a lower dose. In defect detection, a 4.1× increase in sensitivity to relative radial disk shift is observed in the denoised data. Moreover, the LTR-STEM technique may be used to inspect material degradation by tracking changes in diffraction disk intensity, allowing for critical dose estimation and exposure-selective imaging.
\end{abstract}


\end{frontmatter}


\section{Introduction}\label{sec:introduction}

Four-dimensional scanning transmission electron microscopy (4D-STEM) is a powerful technique for quantitative nanoscale materials characterization, where a complete electron diffraction pattern is collected at each scan position \cite{Ophus2019}. It differs from conventional STEM imaging with scintillator-based detectors, in which each scan pixel represents the integrated detector intensity. Different variants of 4D-STEM can be used for tasks such as mapping crystallographic structure, phase, and orientation \cite{Tao2009, Cautaerts2022}, strain fields within materials \cite{Beche2009, Mukherjee2020}, defect characterization \cite{Mills2023, Coupin2023}, electromagnetic field mapping in specimens \cite{Close2015, Krajnak2016}, and other applications. Recent advances in dedicated direct electron detectors have greatly improved the feasibility of conducting high-quality 4D-STEM experiments \cite{Jannis2022, Ercius2024}.

Although frame-based electron detectors, such as monolithic active pixel sensors (MAPS) and hybrid pixel array detectors (HPD or HPAD), provide higher signal-to-noise ratios (SNR) and faster frame rates compared to traditional scintillator-coupled detectors \cite{Faruqi2007}, they are still susceptible to noise. The type of noise depends on numerous factors such as the detector type, acquisition mode, dose rate, and electron energy \cite{Levin2021}. For example, in the common charge integration mode, the accumulated charge in the sensitive layer is read out, which introduces noise from the readout electronics \cite{Faruqi2018}. In MAPS detectors, variations in the energy deposited by individual electrons can produce noise following a Landau distribution \cite{Li2013}. However, both types of noise can be minimized when acquiring very sparse, low-dose diffraction patterns to detect single-electron events in counting mode \cite{Mcmullan2009, Battaglia2009}, effectively reducing the noise to purely shot noise. This form of noise arises from the random fluctuations in the electrons emitted by the source, leading to discrete, stochastic detection events that ultimately set a lower limit on the achievable SNR \cite{Levin2021}.

To reduce the impact of noise in electron diffraction data, especially for 4D-STEM applications, various methods have been developed and used over the years. A simple approach involves smoothing diffraction patterns with filters like Gaussian, non-local means, or Wiener filters \cite{Zhao2023}. While effective and easy to implement, filtering often causes the loss of fine details by removing high-frequency components, restricting their use to specific cases. Another simple method to reduce noise is pixel binning \cite{Zietlow2025}, which boosts the SNR but decreases spatial and reciprocal space information. More robust techniques for 4D-STEM data denoising rely on dimensionality reduction algorithms. These machine learning-based tools decompose multi-dimensional diffraction datasets into linear components, allowing extraction of the most important signals based on variance, while discarding noisier elements. Common tools include principal component analysis (PCA) \cite{Wold1987, Printemps2024}, non-negative matrix factorization (NMF) \cite{Berry2007, Cao2025}, and tensor singular value decomposition (tensor SVD) \cite{Zhang2020}. These are non-local methods, meaning they analyze multiple spatially separated points within the dataset to find common patterns and differentiate signal from noise. The growing popularity of deep learning has also led to neural network approaches for 4D-STEM diffraction data denoising. Unsupervised convolutional neural networks have shown promising results \cite{Sadri2024}, particularly for cases with partial loss of periodicity where decomposition methods may struggle. However, these approaches are computationally demanding and slow, and like decomposition techniques, are inherently non-local.




In this paper, we introduce a method for denoising low- to medium-dose 4D-STEM data by combining electron-event-sparse diffraction-stack acquisition with machine-learning post-processing techniques. At each scan point in the 4D-STEM, multiple event-sparse diffraction patterns are collected at short exposure times around \SI{1}{\micro\second}, which can then be summed to create a higher exposure diffraction pattern. However, before summing, the pattern stack is processed through one of two pipelines to reduce noise, based on electron count sparsity or variance. We demonstrate that these methods significantly decrease noise in diffraction patterns. The denoising approach is local and independent of scan position, making it more flexible than traditional dimensionality reduction techniques, especially for aperiodic data such as that from polycrystalline or disordered materials. Furthermore, the method is computationally efficient and relatively quick, particularly when compared to the leading deep learning methods \cite{Sadri2024}. This new approach, called event-sparse stack denoising or sparse-stack denoising, is illustrated through virtual imaging of polycrystalline gold and defect identification in guanine nanocrystals. Additionally, we show how the acquisition strategy, termed locally time-resolved STEM or LTR-STEM, can be used to estimate beam damage, perform variable exposure diffraction imaging, and determine critical dose thresholds for different materials.


\section{Methods}\label{sec:methods}

\subsection{Samples}\label{sec:methods_samples}

Two samples were analyzed in this study. The first was a Ted Pella Substratek ultrathin metallic support film TEM grid (product no. 21320-25), consisting of a \SIrange{2}{3}{\nano\metre} polycrystalline Au film on a 300 mesh Au grid. Despite its minimal thickness, gold is a strong scatterer, resulting in substantial diffuse background signal in the recorded data. This sample was used to illustrate the impact of denoising on 4D-STEM virtual image quality, with results shown in Section \ref{sec:gold_experimental}.

The second sample consisted of guanine nanocrystals. A dilute suspension of washed crystals was drop-cast onto a 300 mesh Cu TEM grid coated with ultrathin carbon on lacey carbon. This sample was used to demonstrate application-specific denoising for defect identification, with corresponding results presented in Section \ref{sec:results_guanine}.

\subsection{TEM data acquisition and processing}\label{sec:methods_acquisition}

All data presented in this article were recorded on an FEI Talos F200X operated at \SI{200}{\kilo\electronvolt}. The microscope is equipped with a Quantum Detectors MerlinEM 4R direct electron detector for 4D-STEM data acquisition. Although the full detector consists of four chips (512$\times$512 pixels), only a single chip with 256$\times$256 pixels was used for the datasets in this work. Acquisition parameters varied between experiments and are reported in the corresponding sections. A common feature across all datasets is that the diffraction patterns were acquired under sparse-electron conditions, with very short exposure times on the order of \SI{1}{\micro\second}.

\begin{figure}[!ht]
    \centering
    \includegraphics[width=.8\linewidth]{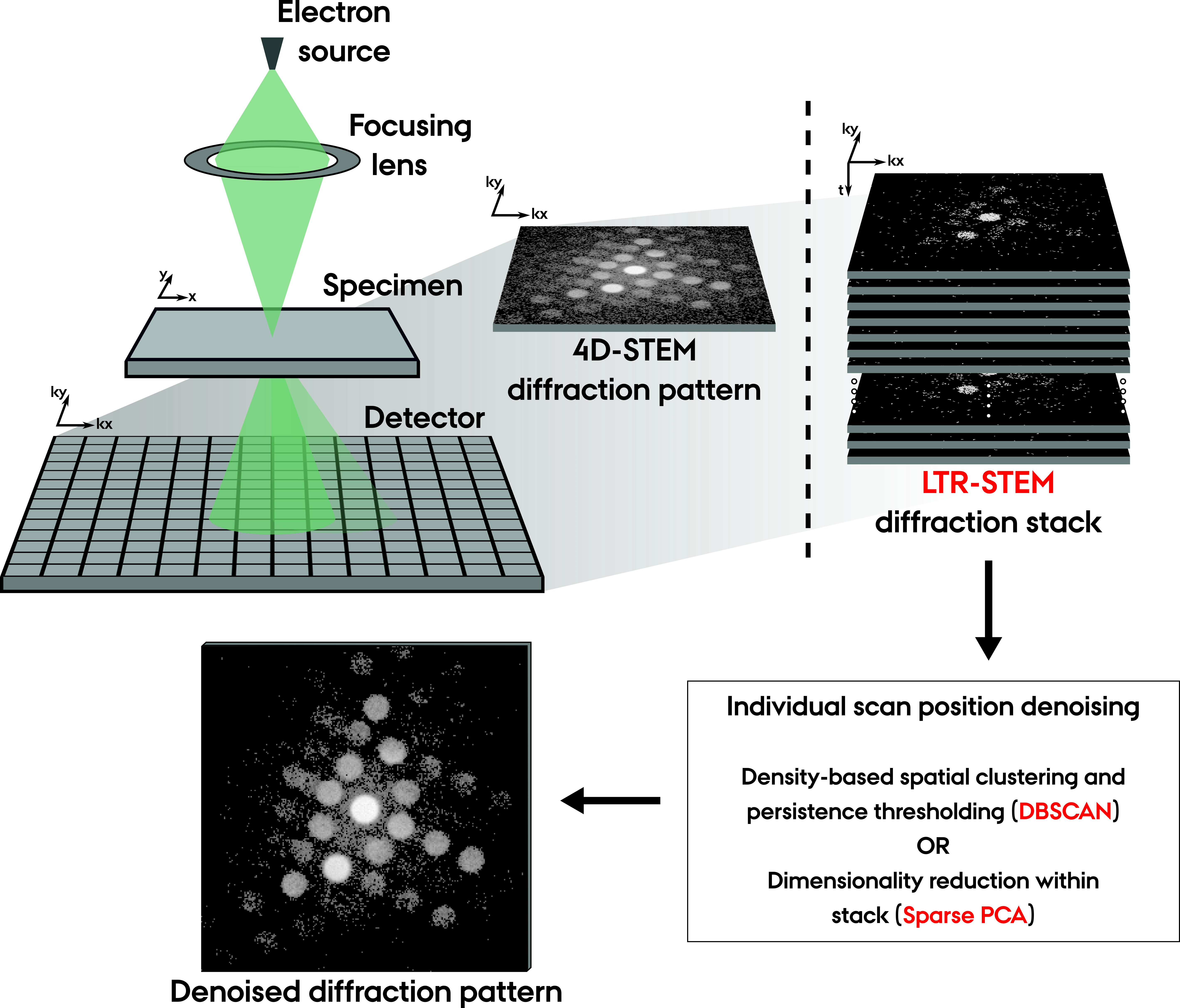}
    \caption{Visualization of the denoising process, showing the acquired diffraction pattern stack (LTR-STEM), the denoising procedures applied to the stack, and the resulting denoised diffraction pattern.}
    \label{fig:ltr_schematic}
\end{figure}

The sparse-stack denoising procedures require that each scan position contains a stack of diffraction patterns. This introduces an additional dimension relative to conventional 4D-STEM, effectively producing a five-dimensional (5D) dataset, here referred to as LTR-STEM, as illustrated in Figure \ref{fig:ltr_schematic}. In practice, this is achieved by adjusting detector parameters while retaining standard 4D-STEM scanning settings. Because a direct electron detector is used, practical constraints related to data throughput must be considered, specifically the rate at which recorded frames can be written to storage. Using the full 256$\times$256 quadrant for diffraction recording makes data transfer the limiting factor for the number of patterns that can be captured per stack at a given number of scan positions. With the current system, we are able to consistently record datasets of shape (64$\times$64$\times$100$\times$256$\times$256), corresponding to 64$\times$64 scan positions, 100 patterns per position, and either 1-bit or 6-bit acquisition depth, both of which are stored internally using 8-bit (1-byte) values. Additional notes on these throughput limitations and potential improvements are provided in Section \ref{sec:results_notes}.

At each scan position, the first diffraction pattern in the stack was found to be unreliable. When reconstructing virtual images, this first frame often introduces a rigid shift relative to subsequent patterns. The most likely explanation is that the detector readout rate exceeds the beam-shift settling time of the scan generator, and under the chosen trigger mode this results in a consistent offset in the first frame. Although several strategies exist to mitigate this effect (e.g., introducing a scan delay or modifying detector trigger timing), in this study we simply discard the first pattern from each stack.

The recorded datasets were converted into HyperSpy signals using the open-source HyperSpy library \cite{hyperspy}, which was also employed for virtual imaging and data visualization. Denoising algorithms were implemented using the open-source scikit-learn library \cite{scikit-learn}. Denoising code, example scripts, and datasets are available in the GitLab repository associated with this article \cite{gitlab_repo}.


\subsection{Simulations}\label{sec:methods_sim}

Multislice simulations of an Au (FCC) nanoparticle oriented along the [001] zone axis were performed using the open-source Python library py\_multislice \cite{pymultislice}. The nanoparticle model, adapted from Barnard (2024) \cite{Barnard2024}, has a diameter of \SI{4}{\nano\meter} and rests on a \SI{1}{\nano\meter} amorphous carbon support film adapted from Ricolleau (2013) \cite{Ricolleau2013}. The full atomic structure used for the simulations is provided in Supplementary Materials Section 1. A probe convergence semi-angle of \SI{3}{\milli\radian} was selected to closely match the experimental conditions used for the polycrystalline Au thin-film dataset presented in Section \ref{sec:gold_experimental}.

For the simulated dataset, a single diffraction pattern was extracted from the center of the symmetric nanoparticle model. This pattern was then divided into a stack of 200 patterns by redistributing the total diffraction intensities randomly but uniformly across the stack. To ensure a clean ground truth for quantitative evaluation, regions of reciprocal space surrounding Bragg reflections were masked to remove diffuse background contributions originating from combined scattering in the gold nanoparticle and the amorphous substrate. Without this masking step, denoising algorithms could appear to surpass the ground truth by artificially suppressing the background intensity between diffraction spots. Finally, Poisson noise with a mean value of $\lambda = 0.007$ was added to the divided and masked pattern stack to emulate low-dose experimental conditions, consistent with those described in Section \ref{sec:gold_experimental}.

\subsection{Denoising algorithms}\label{sec:methods_denoising}



\subsubsection*{Denoising by clustering - DBSCAN}
\label{sec:methods_denoising_dbscan}

The first sparse-stack denoising approach is based on the density-based spatial clustering of applications with noise (DBSCAN) algorithm \cite{Schubert2017}. DBSCAN identifies clusters by searching for groups of points within a radius \textit{eps} that contain at least \textit{min$\_$samples} points \cite{Khan2014}. When applied to diffraction data, \textit{eps} corresponds to the approximate radius of a Bragg reflection, while \textit{min$\_$samples} reflects the local electron count density within a Bragg disk, which is the inverse of the data sparsity. Implementing DBSCAN on sparse diffraction pattern stacks enables the removal of noise by treating isolated intensity counts as outliers.

To improve the robustness of DBSCAN in identifying noisy, low-frequency outliers, a global thresholding step is applied to the entire diffraction stack. After outlier removal is performed independently on each pattern, the remaining counts are accumulated to produce a persistence map. Recurrently occupied pixels across the stack are more likely to represent genuine Bragg reflections. Thresholding is applied in two stages to convert this persistence information into a diffraction mask estimate. In the first stage, a smooth sigmoid function is used to produce a "soft" thresholded pixel map of likely Bragg-reflection locations:
\begin{equation}
\text{mask} = (1+e^{-(pm/N - T)\cdot S})^{-1},
\end{equation}
where $pm$ is the accumulated persistence map, $N$ is the total number of patterns in the stack, $T$ is the threshold level, and $S$ controls the sharpness of the sigmoid transition. Typical threshold values are $T = 0.01$–$0.02$ (i.e., \SIrange{1}{2}{\percent}), with the optimal value depending on noise levels and data sparsity. This soft threshold produces a smooth probability-like mask, which is then multiplied with the DBSCAN-denoised pattern stack. As a final step, a "hard" threshold, or cutoff, is applied to the soft mask to obtain a binary diffraction mask estimate. In practice, the hard threshold is often set equal to the soft threshold parameter. The motivation for using this two-step thresholding procedure is discussed in more detail in Section \ref{sec:results_sim}.


\subsubsection*{Denoising by decomposition - Sparse PCA}
\label{sec:methods_denoising_sparsepca}

The second denoising approach is based on sparse principal component analysis (sparse PCA). Compared with the DBSCAN workflow, this method is simpler to apply because it requires a smaller number of input parameters. PCA belongs to the broader family of dimensionality reduction techniques that decompose datasets into orthogonal components ordered by their explained variance \cite{Jolliffe2016}. The decomposition yields loadings, which describe common features across the dataset, and factors, which indicate where these features occur. In the context of 4D-STEM, the loadings capture the diffraction pattern features with the greatest variability, while the factors identify the scan positions at which these features appear \cite{Bergh2020, Allen2021}. Standard PCA loadings are typically dense (nonzero across all pixels), which reduces interpretability. Sparse PCA addresses this by enforcing sparsity, leading to more localized and physically meaningful component outputs \cite{Zou2006}.

For denoising, the diffraction pattern stack at each scan position is analyzed using sparse PCA, specifying a single output component ($n_{comp} = 1$). The data are then fully reconstructed from this single component. Using more components in the decomposition, or reconstructing the data using fewer components than were extracted, was found to reduce essential diffraction information without improving denoising performance. Conceptually, the single reconstructed pattern functions as a data-driven stack sum in which only pixels exhibiting consistent, structured variance across the stack are retained. Random noise contributions are naturally suppressed and reconstructed with values close to zero. After reconstruction, a final thresholding step is applied at low intensities, typically in the range $10^{-2}$–$10^{0}$ (0.01–1). The workflow therefore depends on only two input parameters: the \textit{alpha} parameter, which controls the sparsity of the sparse PCA components, and the intensity threshold. In practice, the alpha parameter has a negligible effect on denoising performance, thus, the threshold value is the primary parameter to consider.

\section{Results and Discussion}\label{sec:results}


\subsection{Denoising performance on simulated diffraction data}\label{sec:results_sim}

\begin{figure}[!ht]
    \centering
    \includegraphics[width=.95\linewidth]{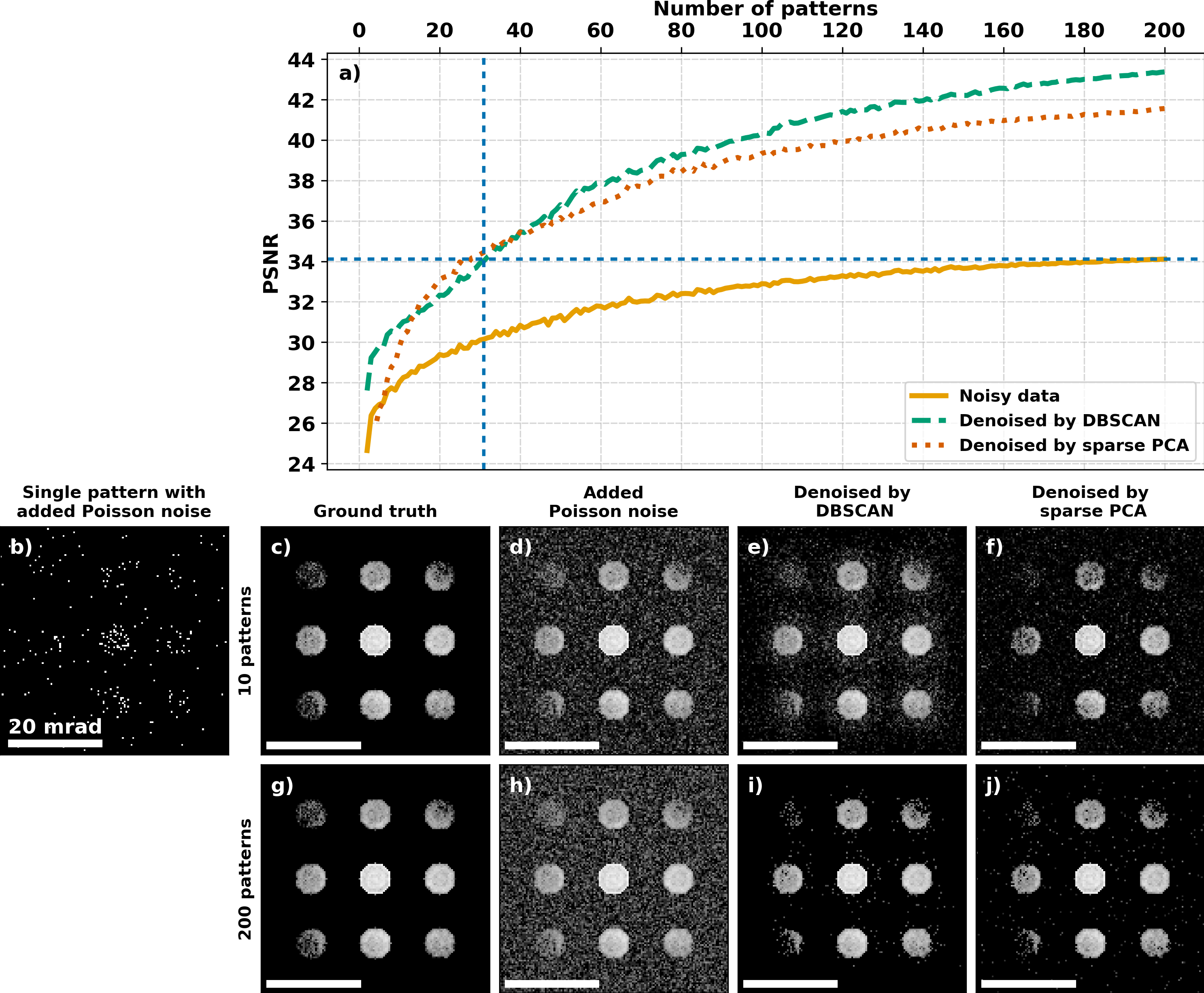}
    \caption{Event-sparse stack denoising of simulated diffraction data with PSNR quantification of the denoised results. a) PSNR curves of noisy data and denoised data obtained using the DBSCAN algorithm (clustering) and the sparse PCA algorithm (decomposition). The blue dashed line marks the point (x = 31) where the PSNR of the clustering-based denoising equals that of the noisy data reconstructed using all 200 components. b) Example diffraction pattern from the stack with added Poisson noise. c,g) Ground truth stack sums before noise addition, using 10 and 200 images, respectively. d,h) Corresponding stack sums with added Poisson noise. e,i) Results of clustering-based denoising. f,j) Results of sparse PCA denoising.}
    \label{fig:psnr_series}
\end{figure}

The simulated Au [001] diffraction pattern was divided into 200 component patterns, masked to remove background around Bragg disks, and corrupted with Poisson noise, as described in Section \ref{sec:methods_sim}. Examples of a noisy component pattern, the ground truth summed pattern, and the noisy summed pattern are shown in Figure \ref{fig:psnr_series}. The full stack was denoised using the DBSCAN and sparse PCA algorithms with identical parameter settings across all tests. For DBSCAN, the parameters were $\text{eps}=9$, $\text{min}_\text{samples}=5$, and a soft–hard threshold of 0.02. For sparse PCA, the threshold parameter was set to 1.

Peak signal-to-noise ratio (PSNR) was used to quantify performance as a function of the number of component patterns $n$ included in the reconstruction (Figure \ref{fig:psnr_series}a). PSNR is defined as \cite{Yang2015}
\begin{equation}
\text{PSNR} = 10 \log_{10}\left(\frac{MAX_I^2}{MSE}\right),
\end{equation}
where $MAX_I$ is the maximum intensity in the ground truth pattern, and $MSE$ is the mean-square error relative to this reference. The latter is the variable term deciding the PSNR.

All PSNR curves exhibit logarithmic growth and converge at large $n$, where noise added to individual components becomes increasingly diluted. For almost the entire range of $n$, both denoising methods yield higher PSNR than the noisy reconstructions. Only at very small $n$ ($<6$) does denoising underperform. DBSCAN generally outperforms sparse PCA, except between $n=11$ and $n=36$. The optimal DBSCAN value $\text{eps}=9$ corresponds to the Bragg disk mask radius used for constructing the ground truth.

As previously mentioned in Section \ref{sec:methods_denoising_dbscan}, a soft–hard double threshold was required for stable DBSCAN performance. Using a single hard cutoff produced discontinuities in the PSNR curve at component numbers proportional to the threshold (e.g., a threshold of 0.02 produces discontinuities at $n={[50,100,150,200]}$, and a threshold of 0.01 at $n={[100,200]}$). Introducing a soft sigmoid threshold suppressed these artifacts and resulted in smooth PSNR behavior.

A notable point occurs at $n=31$, where the PSNR of the DBSCAN-denoised reconstruction matches the PSNR of the noisy reconstruction using all $n=200$ components, marked with a blue dashed line in Figure \ref{fig:psnr_series}a). This suggests that comparable image quality could be achieved using substantially shorter exposures. If one component is acquired at $t=\SI{1}{\micro\second}$, then a conventional diffraction pattern formed by summing 200 components would require $t=\SI{200}{\micro\second}$, whereas the sparse-stack approach requires only $t=\SI{31}{\micro\second}$. The total accumulated dose is therefore reduced to approximately 16\%\footnote{Assuming negligible detector dead time.}, which is advantageous for beam-sensitive materials.



\subsection{Denoising of experimental data for virtual imaging}\label{sec:gold_experimental}

\begin{figure}[!ht]
    \centering
    \includegraphics[width=.95\linewidth]{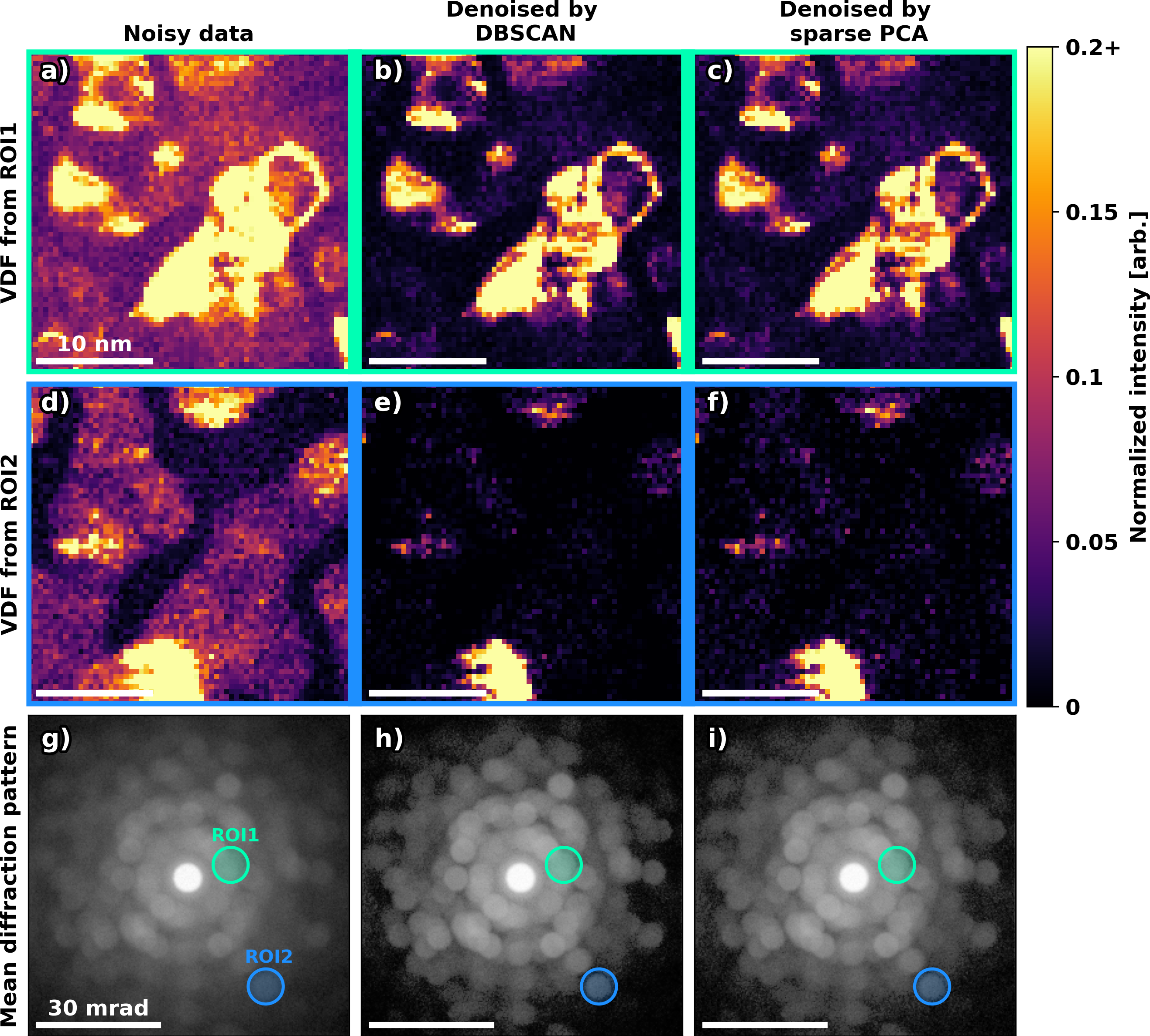}
    \caption{Application of sparse-stack denoising to virtual imaging of nanocrystalline Au. a–c) VDF images from the raw data, clustering-denoised data, and sparse PCA–denoised data using ROI1. d–f) Corresponding VDF images using ROI2. g–i) Mean diffraction patterns with aperture positions indicated.}
    \label{fig:virtual_images}
\end{figure}

Virtual imaging with 4D-STEM data benefits greatly from the application of denoising. To evaluate sparse-stack denoising for virtual imaging, a 4D-STEM dataset was acquired from a \SIrange{2}{3}{\nano\meter} polycrystalline Au thin film (64$\times$64 scan positions, 100 frames per position). Data were recorded with a \SI{3}{\milli\radian} convergence semi-angle, \SI{160}{\milli\meter} camera length, and \SI{0.5}{\micro\second} frame time in 1-bit counting mode. Virtual dark-field (VDF) images reconstructed from two apertures (ROI1 close to the direct beam, and ROI2 at a higher scattering angle) are shown in Figure \ref{fig:virtual_images}. The DBSCAN parameters were $\text{eps}=8$ (approximately equal to Bragg disk radius), $\text{min}_\text{samples}=2$, and soft-hard thresholds of 0.015. The sparse PCA threshold was set to 1.

Despite the small physical thickness of the Au film, gold is a strong scatterer and produces a substantial diffuse background. When a virtual aperture includes regions without well-defined Bragg reflections, this background contributes directly to the VDF images. Denoising suppresses these diffuse counts, leading to improved contrast and a clearer identification of the crystallites contributing intensity to each aperture.

\begin{figure}[!ht]
    \centering
    \includegraphics[width=.95\linewidth]{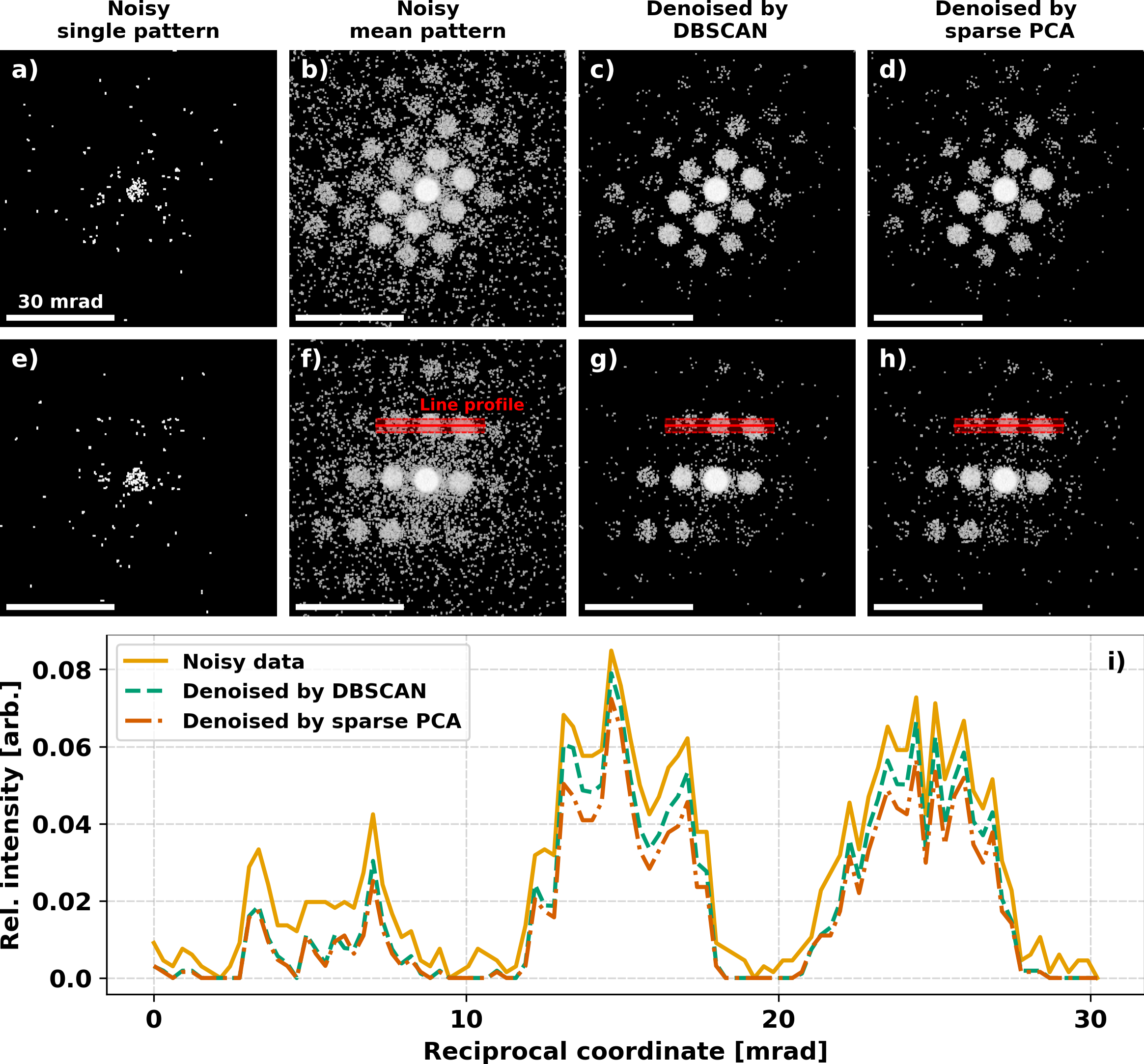}
    \caption{Sparse-stack denoising applied to individual diffraction patterns from the Au dataset. a,e) Single raw patterns. b,f) Summed pattern stacks. c,g) DBSCAN-denoised patterns. d,h) Sparse PCA–denoised patterns. i) Line profiles across a Bragg disk row comparing raw and denoised intensities.}
    \label{fig:single_pos}
\end{figure}

A key limitation of sparse-stack denoising is the partial loss of diffraction disk structure. DBSCAN identifies clusters purely from local electron count density, and sparse PCA retains only components with coherent variance. However, neither method explicitly preserves disk shape. This effect is illustrated in Figure \ref{fig:single_pos}, where two zone-axis patterns from the denoised dataset are compared with their raw counterparts. Line profiles across a row of Bragg disks show reduced intensities for both denoising methods, as seen in Figure \ref{fig:single_pos}i). Some of this reduction reflects correct removal of noise, but information loss becomes unavoidable when the local electron count density within a Bragg disk approaches the noise density targeted for removal. This primarily affects weak, high-angle reflections. In the simulated, denoised data in Figure \ref{fig:psnr_series}, Bragg disks at higher scattering angles often do not retain their full information when denoised as compared to the ground truths.

Between the two methods, DBSCAN retains intensities closer to the raw data, whereas sparse PCA yields more aggressive suppression. DBSCAN parameters can be tuned to mitigate information loss, but typically at the cost of reduced denoising performance. This represents the central trade-off for both denoising approaches.


\subsection{Application-specific denoising for improved defect mapping}
\label{sec:results_guanine}

\begin{figure}[!ht]
\centering
\includegraphics[width=0.9\linewidth]{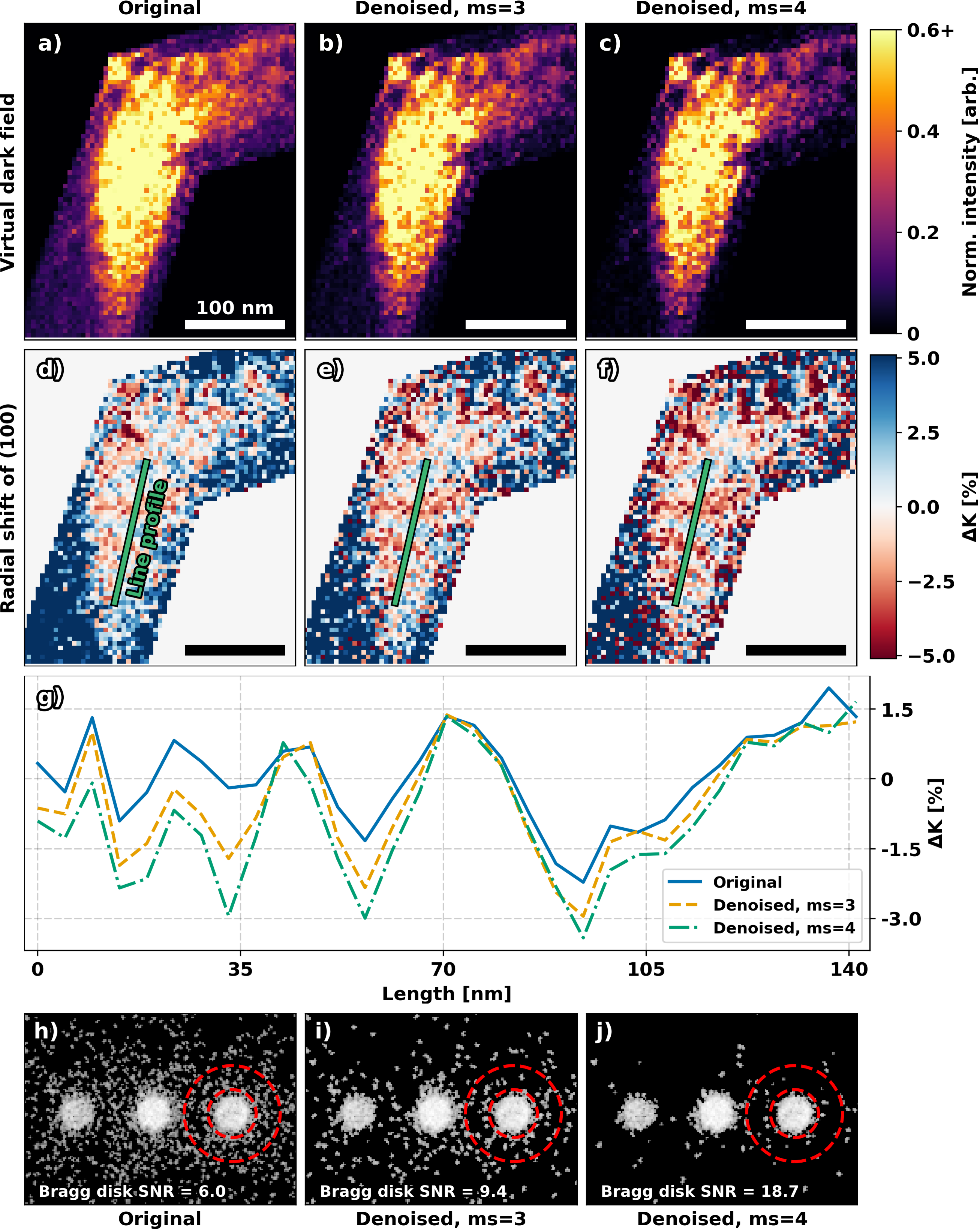}
\caption{Application of sparse-stack denoising to virtual imaging and defect identification in guanine nanocrystallites. a–c) VDF images from the (100) reflection: a) raw data; b–c) denoised results from the DBSCAN algorithm with increasing $\text{min}\_\text{samples}$ values, leading to greater noise reduction. d–f) Corresponding center-of-mass radial disk-shift maps: d) raw data; e–f) denoised results with progressively higher $\text{min}\_\text{samples}$ values. Line profiles across a defect-containing region, extracted from d–f), are plotted in g). h–j) Diffraction patterns from the crystalline region highlighted in a–c), cropped around the (100) reflection row: h) raw data; i–j) denoised results with increasing $\text{min}\_\text{samples}$ values. Diffraction patterns have been rotated by \SI{49}{\degree} counterclockwise for visualization. The (100) reflection used for VDF reconstruction and radial shift measurement is marked by red dotted circles. Inner and outer circles indicate the regions used for SNR calculation, with values reported in h–j).}
\label{fig:guanine_strain_maps}
\end{figure}

Sparse-stack denoising can also be used to enhance defect mapping. As a representative system, we examine guanine nanocrystallites, which commonly exhibit crystallographic bending. A single crystallite oriented near the [001] zone axis is shown in Figure \ref{fig:guanine_strain_maps}. Diffraction data were acquired using a \SI{330}{\milli\meter} camera length, \SI{1}{\micro\second} frame time, 6-bit depth, and a 100$\times$100 scan grid, with 60 diffraction patterns recorded at each position. Virtual images and disk-displacement maps were constructed from the (100) reflection. Disk shifts were quantified by polar transforming each pattern and calculating the center of mass (COM) within a bounded region around the (100) disk \cite{Grieb2021}. While more sensitive to noise than correlation-based methods \cite{Krajnak2016, Pekin2017}, COM provides a straightforward indicator of intensity redistribution.

Denoising was performed with DBSCAN using $\text{eps}=5$, soft-hard thresholds of 0.02, and $\text{min}_\text{samples}$ values between 3 and 4. Although previous analysis suggested optimal performance when $\text{eps}$ matches the Bragg disk radius (approximately 9 pixels), the strongest improvements in disk-shift sensitivity were observed at roughly half that value. This is likely because DBSCAN more effectively captures the intensity redistribution within the disk at this scale caused by, e.g., crystallographic bending. Parameter-dependent comparisons are provided in Supplementary Materials Section 2. Using $\text{min}_\text{samples}=4$ produced a three-fold increase in SNR for the (100) reflection.

Line profiles extracted across the defect-rich region show improved visibility of subtle radial disk shifts after denoising. The raw data exhibits peak-to-trough variations of $\sim$0.9\%. Denoising with $\text{min}_\text{samples}=4$ increases this to $\sim$3.7\%, corresponding to a 4.1× improvement. These variations likely arise from crystallographic bending, consistent with post-scan overview images shown in Supplementary Materials Section 3. Because only the [100] row of reflections is clearly visible in the sparse diffraction data, bending cannot be confirmed solely from the raw patterns. However, sparse-stack denoising makes the associated redistribution of disk intensity significantly more apparent. Raw patterns are provided in Supplementary Materials Section 4.

\subsection{Exposure-selective imaging and critical dose estimation}\label{sec:dose_stuff}

\begin{figure}[!ht]
    \centering
    \includegraphics[width=.95\linewidth]{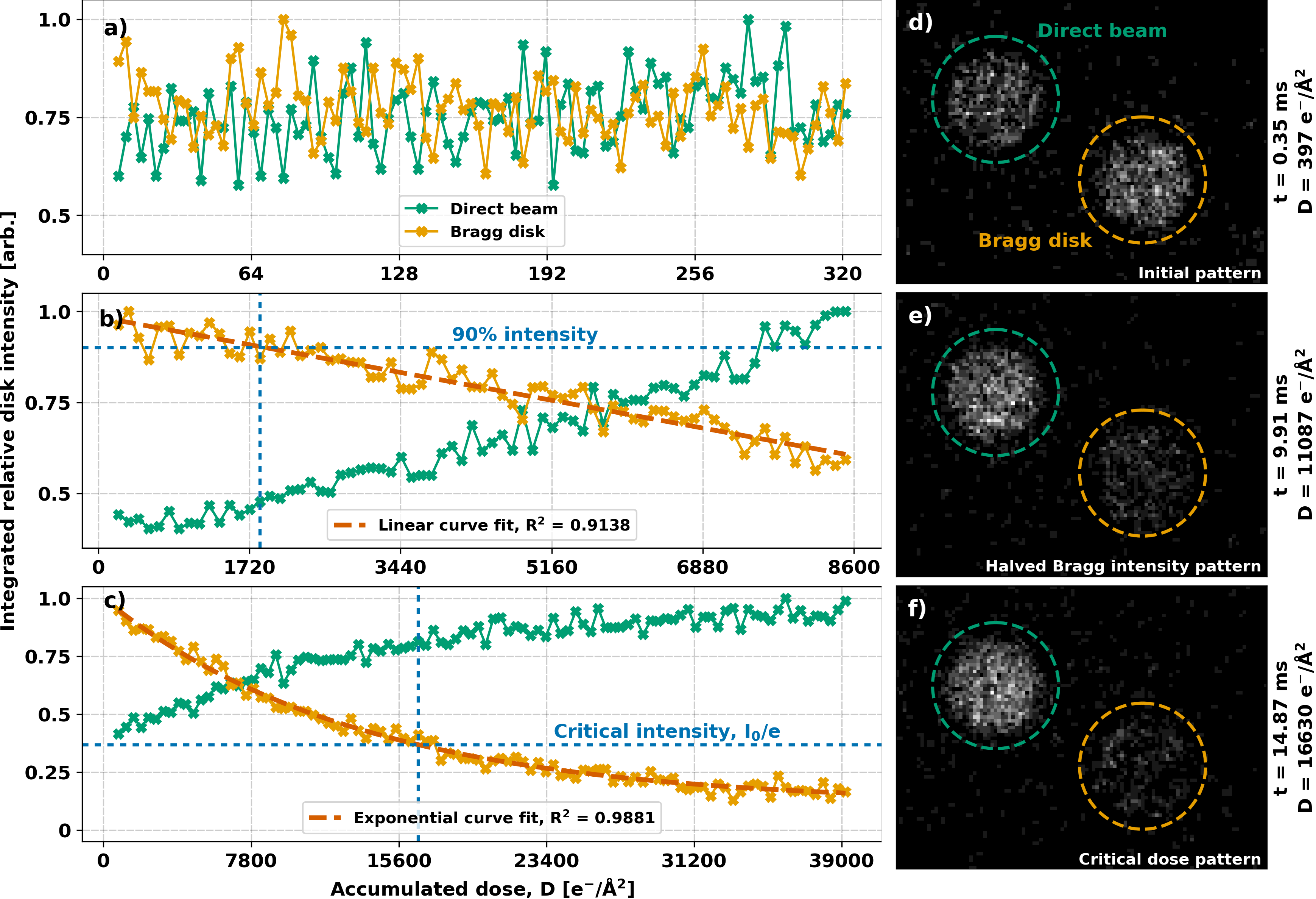}
    \caption{Integrated diffraction disk intensity profiles from the (100)- and (000)-diffraction disks as functions of accumulated electron dose. Operating with the same imaging conditions, but changing condenser C2 aperture size, the profiles in a-c) show the intensity profiles for \SI{20}{\micro\meter}, \SI{50}{\micro\meter}, and \SI{70}{\micro\meter} aperture sizes, respectively. d-f) Direct beam and (100) Bragg disk at different accumulated doses, from the initial pattern, to a reduction to half diffraction disk intensity, and to a diffraction disk intensity reduction to critical dose intensity, respectively. The latter data is from the scan acquired with the \SI{70}{\micro\meter} C2 aperture.}
    \label{fig:intensity_profiles_bragg}
\end{figure}

The guanine dataset used in Figure \ref{fig:guanine_strain_maps} contains 60 patterns per scan position. Inspection of the integrated (100) disk intensity as a function of frame number revealed a monotonic decrease, indicating beam-induced degradation despite the low beam current of \SI{17}{\pico\ampere}. The material is beam-sensitive, and by frame 31, the (100) disk intensity had dropped to 90\% of its initial value. This was the threshold set to avoid structural changes during defect mapping, and all subsequent frames were discarded, effectively performing selective-exposure imaging. A second dataset acquired under identical conditions, shown in Figure \ref{fig:intensity_profiles_bragg}b), reached this 90\% point after only 15 frames, corresponding to an accumulated dose of $\sim$1900 e\textsuperscript{-}/\AA\textsuperscript{2}. Differences between crystallites are presented in Supplementary Materials Section 5, and stem largely from orientation effects. However, all profiles display a nearly linear early-stage decay.

To investigate dose dependence more systematically, several LTR-STEM scans were recorded with identical conditions except for the C2 aperture, which was varied between \SI{20}{\micro\meter}, \SI{50}{\micro\meter}, and \SI{70}{\micro\meter}. These produced beam currents of \SI{3}{\pico\ampere}, \SI{17}{\pico\ampere}, and \SI{30}{\pico\ampere}, respectively.\footnote{Currents were measured using the MerlinEM in counting mode.} Integrated intensities of the (100) disk for each aperture are plotted in Figure \ref{fig:intensity_profiles_bragg}a–c). Diffraction patterns corresponding to progressively higher doses for the \SI{70}{\micro\meter} aperture are shown in Figure \ref{fig:intensity_profiles_bragg}d–f).

Least-squares fitting was performed on the two of the three intensity profiles. The \SI{20}{\micro\meter} dataset remained effectively constant over the full acquisition time ($\sim$\SI{35.4}{\milli\second}\footnote{Although 100 frames acquired at each scan position with a frame time of \SI{1}{\micro\second} would result in a fraction of this duration, the gap time on the MerlinEM even at 1-bit depth acquisition is significantly longer than the \SI{1}{\micro\second} frame time, resulting in very long dwell times.}), indicating negligible beam damage. For the \SI{50}{\micro\meter} dataset, a linear decay provided the best fit (R\textsuperscript{2}$=0.9138$), while the \SI{70}{\micro\meter} dataset followed an exponential decay (R\textsuperscript{2}$=0.9881$), consistent with beam-induced damage kinetics \cite{Chen2020}:
\begin{equation}
I (D) = I_0 \mathrm{e}^{- \frac{D}{D_C}},
\end{equation}
where $D$ the accumulated dose and $D_C$ the critical dose corresponding to an intensity drop by $I_0 / e$. The critical dose is a metric for evaluating significant material degradation, which, for guanine, was determined to be approximately $1.66 \times 10^4$ e\textsuperscript{-}/\AA\textsuperscript{2}. Additionally, a 50\% intensity reduction could be determined at $1.11 \times 10^4$ e\textsuperscript{-}/\AA\textsuperscript{2}.

Orientation-dependent variations introduce uncertainty in the absolute dose response. Because guanine crystallites are often near, but not exactly at, the [001] zone axis, small tilt changes can alter reflection intensities. The LTR-STEM approach mitigates this issue in two ways: (1) only minimal pre-exposure is required to locate suitable regions, and (2) intensity profiles can be averaged over multiple scan positions, reducing noise and site-specific variability. These advantages make LTR-STEM particularly valuable for dose characterization and for \textit{exposure-selective imaging}, in which only early, minimally damaged frames from each scan position are used for analysis.

\subsection{Final notes on sparse-stack denoising}\label{sec:results_notes}

Sparse-stack denoising proved effective across simulated and experimental datasets and for a range of applications. The implemented algorithms are computationally efficient, operate directly on HyperSpy \cite{hyperspy} signals, and support user-defined parameters. Datasets can be loaded lazily using Dask to reduce memory overhead and enable parallel processing \cite{Narayanan2024}. Processing time scales with the number of diffraction components per scan position, with DBSCAN considerably faster than sparse PCA. For the experimental gold dataset in Figures \ref{fig:virtual_images} and \ref{fig:single_pos} (64$\times$64 scan positions, 100 diffraction patterns of 256$\times$256 pixels), DBSCAN required 5 min 41 s, whereas sparse PCA required 86 min 40 s. Both methods exhibit an empirically linear complexity, $O(n)$, with respect to the number of components $n$, as shown in Supplementary Figure S6.

As discussed in Section \ref{sec:methods_acquisition}, data throughput remains a central limitation for LTR-STEM.

In this work, the practical dataset limit was approximately 64$\times$64 scan positions, with 100 diffraction patterns of 256$\times$256 pixels, constrained by the detector frame and gap times. For instance, a frame time of \SI{1}{\micro\second} combined with a \SI{60.7}{\micro\second} gap time (1-bit, 256$\times$256) yields a total of \SI{6.17}{\milli\second}. To ensure stable acquisition across 64$\times$64 scan positions, the dwell time was set to \SI{7}{\milli\second}. Increasing the number of scan positions or stack depth would require proportionally longer dwell times to accommodate the larger data volume. Although not explored here, reducing detector pixel count (e.g., on the MerlinEM) could increase real-space coverage at the cost of reciprocal-space resolution.

At the low beam currents and short exposures used here (\SI{1}{\micro\second}), MerlinEM data are typically $\sim$\SI{99}{\percent} sparse, with the vast majority of pixels containing no counts. Event-based detectors such as TimePix3 avoid storing empty pixels entirely and operate with much shorter effective gap times than frame-based detectors \cite{Jannis2022}. Such detectors therefore offer a promising route to higher throughput for both LTR-STEM acquisition and sparse-stack denoising.



\section{Conclusion}\label{sec:conclusion}

This work introduces a 4D-STEM denoising methodology termed event-sparse stack denoising (or sparse-stack denoising). The approach adds an acquisition dimension by recording multiple event-sparse diffraction patterns at each scan position, an acquisition mode referred to here as LTR-STEM. Two complementary denoising pipelines were developed: a clustering-based method using DBSCAN with global thresholding to remove sparse outliers, and a decomposition-based method using sparse PCA to isolate structured variance within the diffraction stack.

Sparse-stack denoising proved robust across both simulated and experimental datasets. In simulations with Poisson noise, it consistently increased PSNR relative to noisy reconstructions and achieved the performance of a full 200-pattern dataset using only $\sim$16\% of the original exposure. Applied to experimental Au thin films, it improved virtual imaging contrast by suppressing diffuse background. In guanine nanocrystals, it enhanced sensitivity to subtle domain-like defect signatures corresponding to crystallographic bending, yielding a 4.1× improvement in measurable radial Bragg-disk displacements.

The LTR-STEM acquisition strategy also enabled selective-exposure imaging and measurements of dose-limited behavior in beam-sensitive materials. For guanine, monitoring the (100) Bragg disk integrated intensity allowed identification of frame ranges retaining $\geq$90\% of the signal, while discarding the frames where the electron beam had degraded the material. Higher beam currents revealed an exponential decay characteristic of beam damage, from which a critical dose of approximately $1.66 \times 10^4$ e\textsuperscript{-}/\AA\textsuperscript{2} was extracted. Having established this value, subsequent measurements can be performed at lower effective dose rates, reducing damage while preserving interpretability. These results demonstrate that LTR-STEM, combined with sparse-stack denoising, offers a practical pathway for quantifying damage thresholds and optimizing acquisition protocols for beam-sensitive specimens.

Current limitations lie primarily in balancing information preservation against noise removal, particularly for weak high-angle reflections, and in the data throughput constraints imposed by frame-based detectors. Careful parameter tuning mitigates the former, while the latter can be addressed through dwell time adjustments, reduced detector pixel counts, or, more promisingly, event-based detectors. With their negligible gap times and inherently sparse readout, such detectors may remove the primary bottleneck in LTR-STEM acquisition. As these technologies mature, event-sparse stack denoising has strong potential to become a general, scan-position–independent strategy for low-dose 4D-STEM imaging and analysis.

\section*{Acknowledgements}

The authors would like to acknowledge the Danish National Research Foundation (grant no. DNRF189) through the Center for Sustainable Energy Materials (CENSEMAT), and VILLUM FONDEN for the research grant supporting this work (VIL58726). Additionally, big thanks to Henrik Birkedal at the Interdisciplinary Nanoscience Center (iNANO) for supplying us with the guanine samples, and Martin Wibrand Larsen at iNANO for building the atomic gold nanoparticle model and performing multislice simulations for use in this article.


\bibliographystyle{elsarticle-num}
\bibliography{references.bib}

\end{document}